\documentclass[10pt,sigconf,screen,letterpaper]{acmart}

\AtBeginDocument{}

\setcopyright{none}
\renewcommand\footnotetextcopyrightpermission[1]{} 

\acmConference[HotMobile'24]{The 25th International Workshop on Mobile Computing Systems and Applications}{February 28--29, 2024}{San Diego, CA}
\acmPrice{15.00}
\acmISBN{978-1-4503-XXXX-X/18/06}

\acmSubmissionID{5}

\usepackage{amssymb}\usepackage{pifont}\newcommand{\cmark}{\ding{51}}\newcommand{\xmark}{\ding{55}}\usepackage[textfont={normal,bf, it},labelfont=bf]{caption}
\usepackage[labelformat = parens, labelsep = space, textfont=small]{subcaption}

\usepackage{xspace}
\usepackage{enumitem}
\usepackage{tablefootnote}

\definecolor{pro_green}{rgb}{0.0, 0.66, 0.47}

\newcommand{\para }[1]{\medskip \noindent  {\bf #1}}
\newcommand{\1}{{\em (i)}}
\newcommand{\2}{{\em (ii)}}
\newcommand{\3}{{\em (iii)}}

\setlist[itemize]{leftmargin=.12in}

\begin{document}
\sloppy
\title[Designing Spatial Context Awareness for Mobile AR Environment Understanding]{Get-A-Sense: Designing Spatial Context Awareness for Mobile AR Environment Understanding}

\author{Yiqin Zhao}
\orcid{0000-0003-1044-4732}
\affiliation{\institution{Worcester Polytechnic Institute}
 \streetaddress{100 Institute Road}
 \city{Worcester}
 \state{MA}
 \country{USA}
}
\email{yzhao11@wpi.edu}

\author{Ashkan Ganj}
\orcid{0009-0006-3490-0471}
\affiliation{\institution{Worcester Polytechnic Institute}
  \streetaddress{100 Institute Road}
  \city{Worcester}
  \state{MA}
  \country{USA}
}
\email{aganj@wpi.edu}

\author{Tian Guo}
\orcid{0000-0003-0060-2266}
\affiliation{\institution{Worcester Polytechnic Institute}
 \streetaddress{100 Institute Road}
 \city{Worcester}
 \state{MA}
 \country{USA}
 }
\email{tian@wpi.edu}

\renewcommand{\shortauthors}{Zhao et al.}

\begin{abstract}
Physical environment understanding is vital in delivering immersive and interactive mobile augmented reality (AR) user experiences.
Recently, we have witnessed a transition in the design of environment understanding systems, from visual data focused to centering on the concept of spatial context, including user, device, and environment information.
Even though spatial context can benefit the environment understanding tasks, e.g., we demonstrate in a case study that lighting estimation performance can be improved by as much as 59\%, not all environment understanding systems support spatial context. Furthermore, even for the environment understanding systems that support spatial context, not all useful spatial context has been
leveraged; and the design and implementation can differ
vastly. In this paper, we advocate for the design of a spatial context-aware and shared environment understanding system that can effectively support multiple tasks simultaneously.
Fundamentally, there are many practical challenges in designing such a unified environment understanding system. We discuss the challenges in the context of three open questions, \1 how to fully leverage user mobility, \2 how to design a unified data model, and \3 finally how to build a shared system for multiple tasks.
\end{abstract}

\ccsdesc[500]{Computing methodologies~Mixed / augmented reality}
\ccsdesc[500]{Human-centered computing~Ubiquitous and mobile computing systems and tools}

\keywords{Mobile AR; context awareness; environment understanding}

\maketitle

\section{Introduction}
\label{sec:introduction}

Environment understanding is a fundamental task for mobile AR to support immersive and interactive user experiences.
This process typically involves estimating the environment's physical properties, e.g., geometry and lighting, from devices' environment observations.
Taking a representative example, when an AR application user places a virtual object into their physical environment, the AR application requires accurate camera depth information to render the virtual object at the correct distance and generate occlusion effects when needed.

Traditionally, environment understanding systems often leverage the camera sensor inputs and computer vision algorithms to estimate the environment's physical properties.
Over the past decade, as various new sensors are added to AR devices, we have witnessed a growing trend to leverage information extracted from device, user, and environment to improve the accuracy and robustness of environment understanding~\cite{tahara2020retargetable}.
This contextual information, collectively defined as \emph{spatial context}, opens new possibilities of environment understanding across various objectiveness, e.g., lighting, depth, and surfaces.

Spatial context can provide important information to facilitate the environment understanding process by increasing device observations and reducing ambiguities in environment property estimation.
Even though spatial context can be beneficial for environment understanding tasks,
\emph{not all environment understanding systems support spatial context.}
For example, monocular depth estimation uses only visual clues from a single image to estimate the image depth values. Although recent deep models can achieve good accuracy in testing data, many use cases, e.g., temporally-stable depth estimation, in mobile AR are still challenging to these models.

Even for the environment understanding systems that support spatial context~\cite{zhao2022litar,wen2021bundletrack,wvangansbeke_depth_2019}, not all useful spatial context has been leveraged; and the design and implementation can differ vastly.
For example, different systems often use different sets of sensors including IMU, ambient light, and depth.
Moreover, the representations of spatial context data can range from point clouds to signed distance functions, and even nature language.
Such divided data representation discourages spatial context data sharing and exchanging, therefore potentially limiting the overall environment understanding quality.

In this work, we advocate for \emph{a spatial context-aware and multi-task design for environment understanding systems}.
In \S\ref{sec:measurement}, via a case study, we demonstrate the benefits of environment context for one environment understanding task called \emph{lighting estimation} with a recent system~\cite{zhao2022litar}.
To accumulate context information, the system leverages the AR user's mobility and hand movement to capture camera images and other sensory data. To control and reproduce user movements, we use a novel \emph{semi-synthetic simulation} experiment design using the RCareWorld platform~\cite{RCareWorld} which provides photorealistic physical room scan~\cite{Matterport3D} and physically accurate human models~\cite{SMPL-X:2019}.

Specifically, we evaluate three scenarios where spatial context data are discovered from opportunistic environment observations (single and multi-user scenarios), and guided user movements. We observe that spatial context can significantly increase the lighting estimation performance from 13.2db PSNR to 19.7db in multi-user scenarios where context can be shared. However, naively sharing spatial context can incur a high overhead proportional to the number of users. In contrast, guided user movements can achieve better PSNR improvement, 10\% better than multi-user and just 2\% short of the optimal, while only incurring minimal overhead compared to single-user scenarios.

In \S\ref{sec:understanding}, we look into the design of different types of existing environment understanding systems and identify three key open questions, with the goal of designing future AR environment understanding system that is safe and effective.
In particular, we find that existing systems often miss the opportunity to fully exploit the inherent user mobility to obtain more spatial context.
For example, existing works often try to address the challenge associated with the uncontrolled mobility of AR devices, which may introduce incomplete or extreme angles of camera views and, therefore result in poor and inconsistent imaging results~\cite{dehghan2021arkitscenes}.
Rather, we suggest treating mobility as an opportunity and exploring spatial context discovery via guided movement design.

Furthermore, we see that there does not exist a unified data model for spatial context, which limits the flexibility in sharing and coordinating among multiple tasks. Finally, we see that the rapidly increasing environment-understanding capabilities of AR devices usually require real-time and concurrent workload execution~\cite{yi2020heimdall}, which can create conflicts in system resource management~\cite{likamwa2021adaptive} and privacy protection~\cite{farrukh_locin}. As such, we advocate for a shared environment understanding system to deliver services to different tasks. This design is possible because many environment understanding systems already share common components, e.g., sensor access and visual data management; thus developing and running them collectively can be resource-efficient and easier to enforce policies such as privacy management.

We summarize our main contribution as follows:
\begin{itemize}[leftmargin=.12in,topsep=4pt]
    \item We propose a novel simulation-based experiment design with \emph{semi-synthetic data}. This design provides controllable and reproducible AR experimentation with photorealistic rendering of indoor rooms and physical-accurate modeling of human movement.
\item Our measurement study, based on the semi-synthetic simulation, quantifies the impact of spatial context data on a representative environment understanding task, lighting estimation. We use spatial context data collected in three scenarios and show that the lighting estimation performance can be improved by as much as 59\%.
    \item We present a survey on the existing environment understanding system architecture, sensor usage, and data management design. We identify three key open questions towards designing generic system supports for achieving a spatial context-aware and shared environment understanding system.
\end{itemize}

\section{Preliminaries}
\label{sec:preliminaries}

\begin{table}[!t]
	\centering
\caption{Spatial context overview.
        \textnormal{We define three types of spatial context and their corresponding example information.}
    }
    \vspace{-2mm}
	\label{tab:spatial_context_definition}
     \resizebox{0.45\textwidth}{!}{
	\begin{tabular}{@{}l|l@{}}
		\toprule
		\textbf{Type}             & \textbf{Example Information}    \\
		\midrule
		\textbf{User}             & Identity, body pose, eye gaze direction.  \\
		\textbf{Device}           & Sensory data, device model, form factors.  \\
		\textbf{Environment}      & Observations, size measurement, room map.     \\
		\bottomrule
	\end{tabular}
 }
\end{table}

\begin{figure*}[t]
    \centering
\begin{subfigure}[b]{\columnwidth}
        \centering
        \includegraphics[width=.9\linewidth]{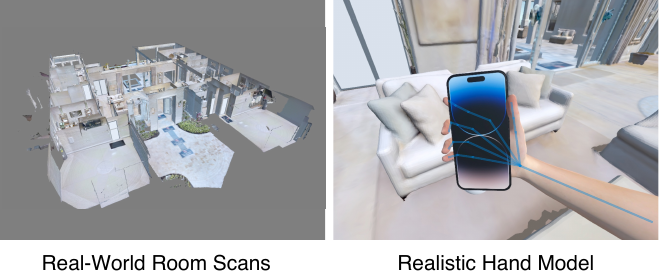}
        \caption{Simulation Setup}
        \label{subfig:ar_system_architecture}
    \end{subfigure}\quad
    \begin{subfigure}[b]{\columnwidth}
        \centering
        \includegraphics[width=.9\linewidth]{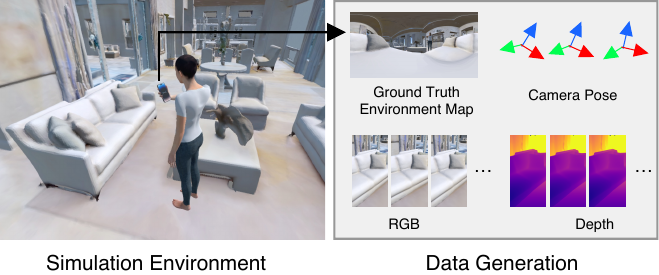}
        \caption{Data Generation}
        \label{subfig:spatial_context_example}
    \end{subfigure}
\vspace{-3mm}
    \caption{
       Overview of our simulation environment semi-synthetic data generation process.
}
    \vspace{-3mm}
    \label{fig:experiment_setup}
\end{figure*}

\para{Environment Understanding in Mobile AR.}
To support immersive and interactive user experiences, AR applications are required to understand several types of environment properties, such as environment lighting, geometry, and surfaces.
Such understanding typically involves the process of estimating the desired properties from camera or sensory data sources.
In traditional AR systems, understanding the environment has been primarily focused on using camera images in favor of the wide availability of camera hardware on AR devices across various factors. Remarkably, by learning from large and diverse datasets, new deep learning models can extract highly accurate information from single or few images as inputs~\cite{birkl2023midas,kirillov2023segany,Jocher_YOLO_by_Ultralytics_2023}.
More recently, mobile AR environment understanding research has emerged to adopt spatial context information during the environment understanding process.
This design opens new opportunities for achieving higher accuracy and robustness in mobile AR environment understanding tasks.

\para{Spatial Context.}
Pioneer research in context-aware AR systems~\cite{starner1998visual} demonstrated that important environment information can be extracted from camera frames for task planning and decision-making.
Such effectiveness proves that important information can be extracted from device environment observations and applied in the process of understanding key properties of the environment.
In recent years, to chase higher accuracy and robustness, new developments of environment understanding systems seek to leverage broad types of information from device, user, and historical camera frames to assist the environment understanding process~\cite{tahara2020retargetable}.
Here we define the term spatial context to describe the available feature information of the user, device, and environment, which all share the same spatial space as the AR application.
In Table~\ref{tab:spatial_context_definition}, we show example information of each corresponding type of spatial context.

Broadly, spatial context data can be collected throughout the AR application usage, as well as from specialized sensors from AR devices or embedded in the environment.
In traditional environment understanding system design, the spatial context information is usually presented in the format of feature point clouds.
Driven by deep learning-based computer vision research, spatial context data can now also be described in more flexible formats, such as implicit functions, embedding vectors, or even natural languages~\cite{wvangansbeke_depth_2019,chen2022text2light,wen2023bundlesdf}.
These new representations open doors to new ways of collecting, managing, and using spatial context in the environment understanding process.

\section{Environment Context Benefits for Lighting Estimation}
\label{sec:measurement}

In this section, we explore the impact of spatial context data through a case study: \emph{lighting estimation with contextual environment observations}.
Lighting estimation refers to the process of estimating omnidirectional environment lighting from limited environment observations, it plays an important role in rendering visually coherent virtual objects.
This task requires an AR system to acquire both in-depth and breadth observations of the room.
During the measurement, we focus on two application scenarios where spatial context data are collected from \1 casual AR application interactions, i.e., opportunistic context collection ($\S$\ref{subsec:opportunistic_context_collection}) under both single- and multi-user scenarios, and \2 guided procedures, i.e., guide context collection ($\S$\ref{subsec:guided_context_collection}).

\subsection{Experiment Design}
\label{subsec:experimental_setup}

\para{The AR Virtual Object Placement Task.}
To test the AR system's performance in real-world scenarios, we perform \emph{the AR virtual object placement task}, which represents one of the most common use cases of mobile AR, e.g., virtual furniture shopping.
The workflow of this task involves a participant using an AR application to place a virtual object on a physical environment surface, e.g., a room floor, and move around their AR device to observe the object.
During the process, we require the participants to only move their hands and arms and keep the AR device looking at the virtual object placement position from different angles.
This movement simulates the "look-around" patterns in the AR application.
This task represents a challenging environment understanding scenario where the AR systems are required to leverage the device mobility to discover the most environment information to deliver a plausible experience.

\para{Semi-Synthetic Data.}
Achieving precise and controllable AR experiments has been a long-standing challenge.
To address the difficulties in minimizing the synthetic-to-real gaps, we present a novel approach to facilitate the process by using \emph{a semi-synthetic simulation-based experiment environment}.
Specifically, our experimentation design differs from traditional simulation-based experiments in two key respects:
\1 \emph{photorealistic rendering} and \2 \emph{physically-accurate human modeling}.
First, we set up the simulation environment with the RCareWorld~\cite{RCareWorld} platform and high-fidelity assets from Matterport3D~\cite{Matterport3D}.
The scanned 3D rooms are used to simulate the indoor environment where our experiments will take place.
Next, to produce a realistic simulation of device camera movements, we use the human avatars RCareWorld~\cite{RCareWorld}, a human-centric simulation environment where the human avatar joints are derived from clinical data.
Last but not least, we set the virtual AR device to use a real-world mobile phone (iPhone 14 Pro) form factors and camera parameters.

To generate our experiment data, we manually selected 12 different environment positions in the simulated room to execute the virtual object placement task.
For each placement position, we apply an animation on the avatar's wrist joint to simulate the \emph{look-around} moving pattern.
We illustrate the data generation process in Figure~\ref{fig:experiment_setup}.
Throughout the entire interaction, we extracted the camera pose, RGB, and depth images.
To allow quantitative evaluation of the lighting estimation environment understanding task, we have also extracted the ground truth environment map at the placement position using a virtual panoramic camera in Unity.
To simulate the multi-user scenario, we manually set each engaged user to stand around the placement position in a circular formation.
In total, we generated 36 sets of experiment data, where each set contains 150 frames of camera pose, RGB, and depth images.
Combining photorealistic rendering and physically accurate human modeling, we believe our simulation-based experiment represents a step forward to flexible and controllable AR experiment design.

\begin{figure}[t]
    \centering
\begin{subfigure}[b]{0.45\columnwidth}
        \centering
        \includegraphics[width=\linewidth]{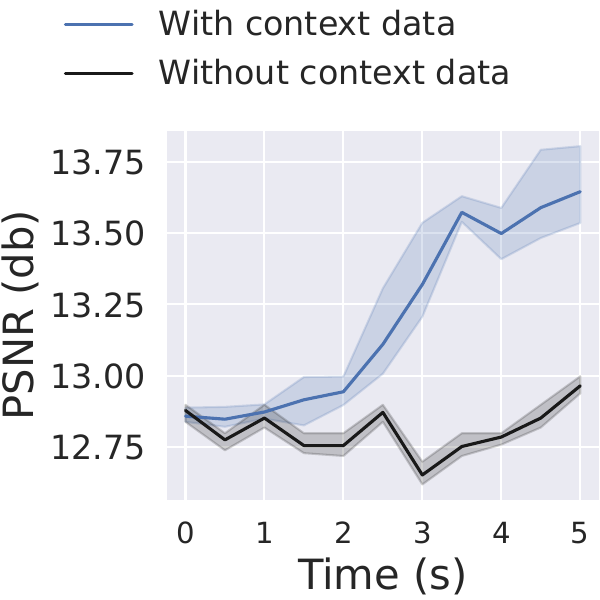}
        \caption{Single User}
        \label{subfig:opportunistic_single_user}
    \end{subfigure}\quad
    \begin{subfigure}[b]{0.45\columnwidth}
        \centering
        \includegraphics[width=\linewidth]{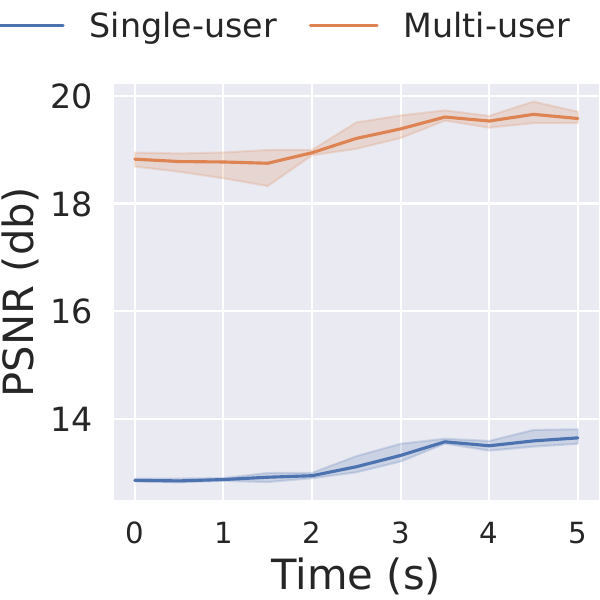}
        \caption{Multi User}
        \label{subfig:opportunistic_multi_user}
    \end{subfigure}
\vspace{-3mm}
    \caption{The impact of contextual environment observation.
        \textnormal{We show that accumulating environment observations over time pose positive but limited feedback on the lighting estimation performance. In the multi-user scenario, inter-user observation sharing allows significant lighting estimation performance improvement.}
    }
    \vspace{-3mm}
    \label{fig:opportunistic_context}
\end{figure}

\subsection{Opportunistic Context Collection}
\label{subsec:opportunistic_context_collection}

\para{Single-user Scenario.}
We first measure the contextual environment observation impact on lighting estimation in the single-user scenario when executing the object placement task.
For each selected placement position, we use the generated camera RGB, depth, and tracking data to simulate the sensory data stream available to the AR device.
We removed LitAR's environment observation constraints, i.e., the number of environment observations to keep, to fully test its capabilities of lighting estimation.
Figure~\ref{subfig:opportunistic_single_user} shows the LitAR-generated environment map PSNR changes during the application time.
On average, LitAR with spatial context outperforms the baseline, where only the current camera frame is used (i.e., without context data), but by only $0.9$db PSNR.
This is largely due to the imperfect context data capture process, i.e., only relying on the user's hand movements.
Upon manually inspecting the camera images and movements, we found that although the virtual object surrounding environment is clearly captured by the AR device camera, the user's hand movements only induce overlapped views, resulting in only a small amount of observation coverage increase.

\begin{figure}[t]
\centering
    \centering
\begin{subfigure}[b]{0.45\columnwidth}
        \centering
        \includegraphics[width=\linewidth]{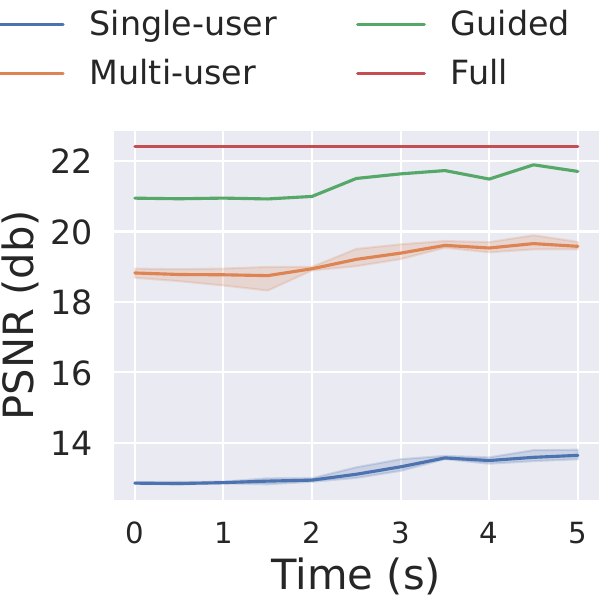}
        \caption{Guided}
        \label{subfig:guided}
    \end{subfigure}\quad
    \begin{subfigure}[b]{0.45\columnwidth}
        \centering
        \includegraphics[width=\linewidth]{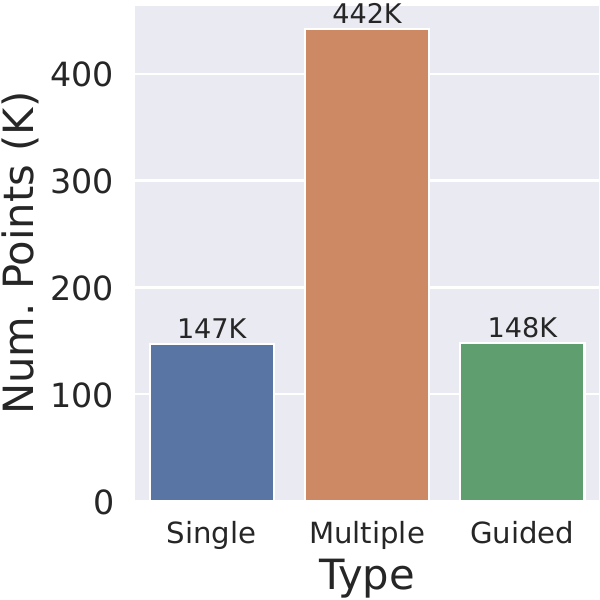}
        \caption{Data Size}
        \label{subfig:context_data_size}
    \end{subfigure}
\vspace{-3mm}
    \caption{The impact of guided context collection.
        \textnormal{Guided context collection outperforms multi-user context sharing by 10\% while using only 33\% of the memory.}
    }
    \vspace{-5mm}
    \label{fig:adversial_robustness}
\end{figure}

\para{Multi-user Scenario.}
Next, we test the multi-user scenario where three users are engaged in the virtual object placement task and share environment observations with each other.
In this scenario, contextual environment observations are collected from user movement and multi-user sharing.
Figure~\ref{subfig:opportunistic_multi_user} compares the lighting estimation performance.
We observe that by sharing contextual environment observations from multiple users, the lighting estimation performance significantly improves by an average of 40\%, compared to the single-user scenario.
This is because the multi-user scenario can capture context information from different angles and thus lead to higher coverage of the environment.
In practice, we believe the multi-user application scenario can be extrapolated to more generalized multi-source context collection cases, such as edge IoT-assisted AR, multi-application context sharing, and multi-time context sharing.

\begin{table*}[!t]
	\centering
\caption{A survey of context-aware environment understanding system design.
        \textnormal{}
}
    \vspace{-3mm}
	\label{tab:context_survey}
    \resizebox{0.78\textwidth}{!}{
	\begin{tabular}{@{}l|rrrc@{}}
		\toprule
		                                                   & \textbf{Task} & \textbf{Context Data Representation} & \textbf{Sensor Usage}   & \textbf{Context Refinement} \\
		\midrule
		\textbf{Text2Light~\cite{chen2022text2light}}      & Lighting Estimation   & Nature Language                      & N/A             					& \xmark  \\
		\textbf{LitAR~\cite{zhao2022litar}}                & Lighting Estimation   & Dense and Sparse Point Cloud         & IMU, Ambient Light, Depth    		& \xmark  \\
		\textbf{BundleTrack~\cite{wen2021bundletrack}}     & Object Tracking       & Sparse Point Cloud                   & Depth             					& \cmark  \\
		\textbf{BundleSDF~\cite{wen2023bundlesdf}}         & Object Tracking       & Signed Distance Function             & Depth            					& \cmark  \\
		\textbf{Sparse SPN~\cite{wvangansbeke_depth_2019}} & Depth Estimation      & Sparse Point Cloud                   & Depth             					& \xmark  \\
		\textbf{Sartipi et al.~\cite{sartipi2020deep}}     & Depth Estimation      & Sparse Point Cloud                   & IMU             					& \cmark  \\
		\bottomrule
	\end{tabular}
 }
\end{table*}

\subsection{Guided Context Collection}
\label{subsec:guided_context_collection}

Finally, we test the impact of spatial context data collected by guided user movement.
Before executing the virtual object placement task, we collect environment observation data using the bootstrapping movement pattern introduced in LitAR.
During the guided movement, LitAR collects large coverage environment observations and converts them into sparse point clouds.
Figure~\ref{subfig:guided} compares the lighting estimation performance difference.
We see that guided context collection leads to better performance when compared to opportunistic context collection. For example, it outperforms multi-user context sharing by 10\%.
Additionally, when comparing the performance of using full environment observations, guided context collection has only 0.5db lower PSNR.
Furthermore, guided context collection only incurs 33\% of the memory usage when compared to the multi-user scenario.
This suggests the need for using a well-designed context-capturing process rather than just naively relying on multi-user sharing.

\section{Open Questions and Inspirations}
\label{sec:understanding}

\begin{figure}[t]
    \centering
    \includegraphics[width=.45\textwidth]{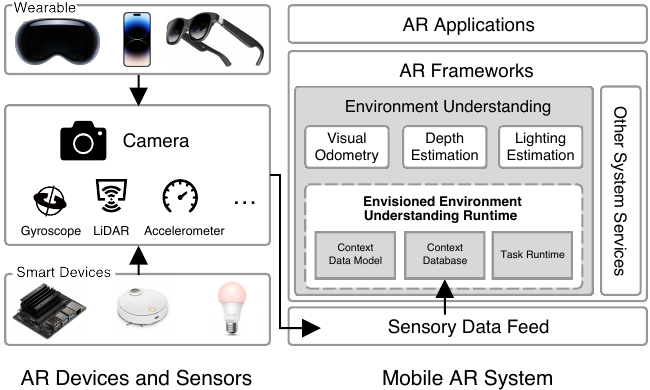}
    \vspace{-3mm}
    \caption{
       Envisioned environment understanding system.
       \textnormal{AR systems collect various types of sensor data, e.g., from AR and smart devices. We envision future AR environment understanding systems to be built on runtime with centralized spatial context data sharing and management.
}
}
    \vspace{-5mm}
    \label{fig:context_overview}
\end{figure}

Following the experimenting results of lighting estimation in the AR virtual object placement task, in this section, we discuss the open questions and inspirations towards achieving spatial context awareness in the general designs of mobile AR environment understanding systems.
Figure~\ref{fig:context_overview} depicts the interaction between the environment understanding system and the rest of the AR ecosystem, as well as what our envisioned spatial context-aware system looks like.

\vspace{-2mm}
\subsection{Better Spatial Context Discovery}
\vspace{-2mm}

\emph{--- How can we fully leverage user mobility to discover more spatial context?}

As a distinctive feature of mobile AR, user mobility naturally introduces continuous camera movements throughout the AR applications.
Such mobility can be leveraged to obtain more observations of the environment, therefore enabling the potential to increase spatial context discovery.
Several systems have been designed to leverage the continuous camera observations from users' mobility for various environment-understanding tasks, e.g., lighting estimation~\cite{zhao2022litar}, SLAM~\cite{mulloni2013init}, and plane detection~\cite{nowacki2020capabilities}, etc.

In previous experiments, we have learned that opportunistically discovering spatial context can be insufficient for the lighting estimation task.
On the other hand, meticulously designed guided context discovery has proven to be an effective and efficient context discovery method for lighting estimation.
Such user engagement has a sweet point, where only minimal user efforts are required but rich spatial context can be discovered.
While the guided context discovery sheds light on designing future AR environment understanding systems, challenges lie in many aspects of approaching this sweet point.
In particular, online evaluation metrics of user movement need to be developed to estimate the performance impact of the discovered spatial context, and interactive feedback should be given to guide users to ease the process.

\subsection{Easier Spatial Context Management}

\emph{--- How can we design a unified spatial context data model?}

Environment understanding is fundamentally a data-driven process, and the spatial context data is key in the overall system.
However, existing environment-understanding systems implement the data processing with significant differences, as shown in Table~\ref{tab:context_survey}.
For example, spatial context data can exist in various representations, including point clouds, signed distance functions, and even nature language.
Such divided data representation discourages spatial context data sharing and exchanging, potentially limiting the overall environment understanding quality.

The challenges in unifying context data models lie in building standard data structures and exchanging protocols for spatial context data providers, e.g., different camera models, and consumers, e.g., different environment understanding systems.
While many proposals were made for context data structure, standardization has yet to be achieved.
For example, in the ARKit\footnote{https://developer.apple.com/augmented-reality/} definition, environment geometries are represented as meshes, and key entities, e.g., tracked objects, are represented as anchors.
Although these data structures can be accessed programmatically, their design is oriented to specific context data providers and consumers.
Such design limits the potential of connecting individual AR devices into networked AR systems, where important environment information can be shared and exchanged.

\subsection{More Efficient Spatial Context Service}

\emph{--- How can we embed spatial context awareness into the environment-understanding task runtime systems?}

Mobile AR systems support environment understanding by providing high-level abstractions of device hardware and software services.
When useful information arrives at environment understanding systems, copies of the information will be in their own memory space.
This self-autonomy design, however, leads to inefficient context data management, as multiple copies of the environment information could be kept in different environment understanding systems.
Moreover, recent work has shown that mismanaging spatial context data can result in information leakage~\cite{farrukh_locin} which threatens users' privacy.
As environment-understanding tasks are being increasingly added to AR systems, the need arises for a standardized supporting system, i.e., environment-understanding task runtime.

We believe that, agnostic to the computations involved in the environment understanding process, the storage and management of spatial context data can be separated from existing systems architectures and managed as a real-time database inside the runtime.
This centralized data management policy can provide first-party design from mobile AR system developers to achieve better system resource efficiency and privacy protection.
Moreover, we have noticed that some environment understanding systems allow context data to be refined over time, as shown in Table~\ref{tab:context_survey}, when environment observations change.
With centralized spatial context data management, the refinement can propagate to more than one environment understanding task, leading to better overall performance.
In addition, we believe having a standardized environment understanding runtime system can also encourage developers and researchers to \1 build new algorithms to fully leverage the spatial context data and \2 capture new datasets with these spatial context data.

\section{Conclusion and Future Work}
\label{sec:conclusion}

In this paper, we quantified the benefits of spatial context on environment understanding tasks via a case study using a semi-synthetic simulation-based experiment. We found that high-quality environment context, from multi-user and guided movement, can improve lighting estimation task performance by 10\% and 59\%. However, designing a spatial context-aware and multi-task environment understanding system can be challenging, as showcased by our analysis of existing systems. We identified three key design questions and inspirations for better discovering spatial context, facilitating coordination and information sharing among tasks, and providing spatial context-aware service.
As part of future work, we plan to enhance the simulator by adding human muscle models to better simulate user movement and thus quantify different spatial context benefits.

\scriptsize{
\bibliographystyle{abbrv}
\bibliography{main}
}

\end{document}